\documentclass[useAMS,usenatbib]{mn2e}
\usepackage{natbib}
\usepackage{epsf}
\usepackage{graphicx}
\usepackage{textcomp}

\def\simlt{\mathrel{\rlap{\lower 3pt\hbox{$\sim$}}
        \raise 2.0pt\hbox{$<$}}}
\def\simgt{\mathrel{\rlap{\lower 3pt\hbox{$\sim$}}
        \raise 2.0pt\hbox{$>$}}}
\def\1752{XTE J1752--223}

\title[A jet rebrightening in XTE J1752--223]{A late jet rebrightening revealed from multi-wavelength monitoring of the black hole candidate XTE J1752--223\thanks{Based on observations collected at the European Southern Observatory, Chile, under ESO Programme ID 086.D-0610, the William Herschel Telescope operated on the island of La Palma by the Isaac Newton Group in the Spanish Observatorio del Roque de los Muchachos of the Instituto de Astrof\'isica de Canarias and the Faulkes Telescopes at Haleakala, Maui, USA and Siding Spring, Australia.}}
\author[D.M. Russell et al.]{D.M. Russell$^{1}$\thanks{E-mail: d.m.russell@uva.nl}, P.A. Curran$^{2}$, T. Mu\~noz-Darias$^{3}$, F. Lewis$^{4,5}$, S. Motta$^{3}$,
\newauthor
H. Stiele$^{3}$, T. Belloni$^{3}$, J.C.A. Miller-Jones$^{6}$, P.G. Jonker$^{7,8,9}$, K. O'Brien$^{10}$,
\newauthor
J. Homan$^{11}$, P. Casella$^{12}$, P. Gandhi$^{13}$, P. Soleri$^{14}$, S. Markoff$^{1}$, D. Maitra$^{15}$,
\newauthor
E. Gallo$^{15}$, M. Cadolle Bel$^{16}$
\\
$^{1}$Astronomical Institute `Anton Pannekoek', University of Amsterdam, P.O. Box 94249, 1090 GE Amsterdam, the Netherlands\\
$^{2}$Laboratoire AIM, CEA/IRFU-Universit\'e Paris Diderot-CNRS/INSU, CEA DSM/IRFU/SAp, Centre de Saclay, F-91191  \\
Gif-sur-Yvette, France\\
$^{3}$INAF-Osservatorio Astronomico di Brera, Via E. Bianchi 46, I-23807 Merate (LC), Italy\\
$^{4}$Faulkes Telescope Project, University of Glamorgan, Pontypridd CF37 1DL, UK\\
$^{5}$Department of Physics and Astronomy, The Open University, Walton Hall, Milton Keynes MK7 6AA, UK\\
$^{6}$International Centre for Radio Astronomy Research - Curtin University, GPO Box U1987, Perth, WA 6845, Australia\\
$^{7}$SRON, Netherlands Institute for Space Research, Sorbonnelaan 2, 3584 CA, Utrecht, the Netherlands\\
$^{8}$Harvard-Smithsonian Center for Astrophysics, 60 Garden Street, Cambridge, MA 02138, USA\\
$^{9}$Department of Astrophysics, IMAPP, Radboud University Nijmegen, PO Box 9010, NL-6500 GL Nijmegen, the Netherlands\\
$^{10}$Department of Physics, University of California, Santa Barbara, CA, USA\\
$^{11}$MIT Kavli Institute for Astrophysics and Space Research, 70 Vassar Street, Cambridge, MA 02139, USA\\
$^{12}$School of Physics and Astronomy, University of Southampton, Southampton, Hampshire SO17 1BJ, UK\\
$^{13}$ISAS, Japan Aerospace Exploration Agency, 3-1-1 Yoshinodai, chuo-ku, Sagamihara, Kanagawa 229-8510, Japan\\
$^{14}$Kapteyn Astronomical Institute, University of Groningen, PO Box 800, 9700 AV Groningen, the Netherlands\\
$^{15}$Department of Astronomy, University of Michigan, 500 Church Street, Ann Arbor, MI 48109, USA\\
$^{16}$ESAC, ISOC, Villanueva de la Cañada, Madrid, Spain\\
}
\begin{document}


\pagerange{\pageref{firstpage}--\pageref{lastpage}} \pubyear{2011}

\maketitle

\label{firstpage}

\begin{abstract}
We present optical monitoring of the black hole candidate \1752 during its 2009--2010 outburst and decay to quiescence. The optical light curve can be described by an exponential decay followed by a plateau, then a more rapid fade towards quiescence. The plateau appears to be due to an extra component of optical emission that brightens and then fades over $\sim 40$ days. We show evidence for the origin of this optical `flare' to be the synchrotron jet during the decaying hard state, and we identify and isolate both disc and jet components in the spectral energy distributions. The optical flare has the same morphology and amplitude as a contemporaneous X-ray rebrightening. This suggests a common origin, but no firm conclusions can be made favouring or disfavouring the jet producing the X-ray flare. The quiescent optical magnitudes are B $\geq 20.6$, V $\geq 21.1$, R $\geq 19.5$, i$^{\prime}$ $\geq 19.2$. From the optical outburst amplitude we estimate a likely orbital period of $< 22$ h. We also present near-infrared (NIR) photometry and polarimetry and rare mid-infrared imaging (8 -- 12 $\mu$m) when the source is nearing quiescence. The fading jet component, and possibly the companion star may contribute to the NIR flux. We derive deep mid-IR flux upper limits and NIR linear polarization upper limits. With the inclusion of radio data, we measure an almost flat jet spectral index between radio and optical; $F_\nu \propto \nu^{\sim +0.05}$. The data favour the jet break to optically thin emission to reside in the infrared, but may shift to frequencies as high as the optical or UV during the peak of the flare.
\end{abstract}


\section{Introduction}

Since the first detections of X-ray transients in the 1960s, monitoring the X-ray evolution of their outbursts has been an active field of research \citep*[e.g.][]{evanet70,chenet97}. Now it is known that multi-wavelength studies, from radio to $\gamma$-ray, over various timescales from milliseconds to decades, can help to explain these objects in detail. Contemporaneous X-ray/radio monitoring of black hole X-ray binary (BHXB) outbursts have revealed correlated behaviour which firmly links the properties of the inflowing matter with those of the outflow in the form of jets \citep*[e.g.][]{tanaet72,hannet98,corbet00,corbet03,gallet03,kordet06,kaleet06}. The optical--infrared (OIR) emission from the outer accretion disc and the jet are also correlated with the X-ray properties on long (weeks--months) and short (less than seconds) timescales \citep*[e.g.][]{motcet85,eikeet98,russet06,caseet10}.

In recent years a global picture has been developed \citep*{fendet04,fendet09} in which the X-ray luminosity, hardness and history are indicative of the radio characteristics. When the X-ray spectrum is hard (the `hard state'; historically the `low/hard state'), a steady, compact jet is launched, which has a flat-spectrum ($\alpha \approx 0$ where $F_{\nu} \propto \nu^{\alpha}$) from radio to infrared frequencies produced by optically thick, self-absorbed synchrotron emission. At some higher frequency, possibly in the OIR domain \citep*{market01,market03,heinsu03}, this emission breaks to an optically thin spectrum, with $\alpha \approx -0.6$. The break defines the minimum total energy contained in the jet, but its location in the spectrum has only been directly observed or inferred to exist in the infrared in a few BHXBs \citep{corbfe02,coriet09,rahoet11,gandet11} and is uncertain in most sources.

The jet appears to be quenched at both radio and OIR regimes when the X-ray spectrum is soft \citep[the soft, or `high/soft' X-ray state; e.g.][]{gallet03,homaet05,fendet09}. In the last few years it has been shown that as well as the radio behaviour, the OIR properties (flux and colour) of some sources can now also be predicted from the X-ray properties \citep[and vice versa;][]{coriet09,russet11b}. These correlations between the different components of the system help us to understand the process of accretion onto compact objects and jet formation in BHXBs.

\1752 is an X-ray transient discovered by the Rossi X-ray Timing Explorer (RXTE) and \textit{Swift} satellites in 2009 October \citep{market09}. The outburst lasted $\sim 8$ months and was observed by the Monitor of All-sky X-ray Image (MAXI) and RXTE satellites \citep{nakaet10,shapet10} at X-ray energies, \textit{Swift} \citep[][hereafter C11]{curret11} at X-ray and optical/UV \citep[the optical counterpart was discovered by][]{torret09a}, Faulkes Telescope and SMARTS at OIR \citep{russet10b,kaleet11} and the Australia Telescope Compact Array (ATCA), the European VLBI Network (EVN), the Very Long Baseline Array (VLBA) and the Expanded Very Large Array (EVLA) at radio frequencies \citep[][Jonker et al. in preparation]{brocet09,yanget10,millet11}. The source evolved from a hard X-ray state to a soft state and back to a hard state at a lower luminosity before fading towards quiescence; overall showing a behaviour typical for a BHXB outburst \citep{fendet04,bell10}. A mildly relativistic radio jet \citep{yanget10,millet11} was also reported. The source was suggested to be a black hole candidate due to its state evolution and its state dependent X-ray fast timing properties \citep{munoet10,shapet10}.
Here we present multi-waveband optical monitoring during the decay of the outburst of \1752 towards quiescence. Together with X-ray, IR and radio observations, we constrain the contribution and spectrum of the accretion disc and the jet to these wavelengths and follow their evolution throughout the outburst decay.
In Section 2 we describe the data and its reduction. The multi-wavelength evolution of the outburst is analyzed in Section 3, in which we construct spectral energy distributions (SEDs), attempt to isolate disc and jet emission, test for optical--X-ray correlations and discuss the results. We summarize our findings in Section 4.

\begin{figure}
\centering
\includegraphics[width=8.3cm,angle=0]{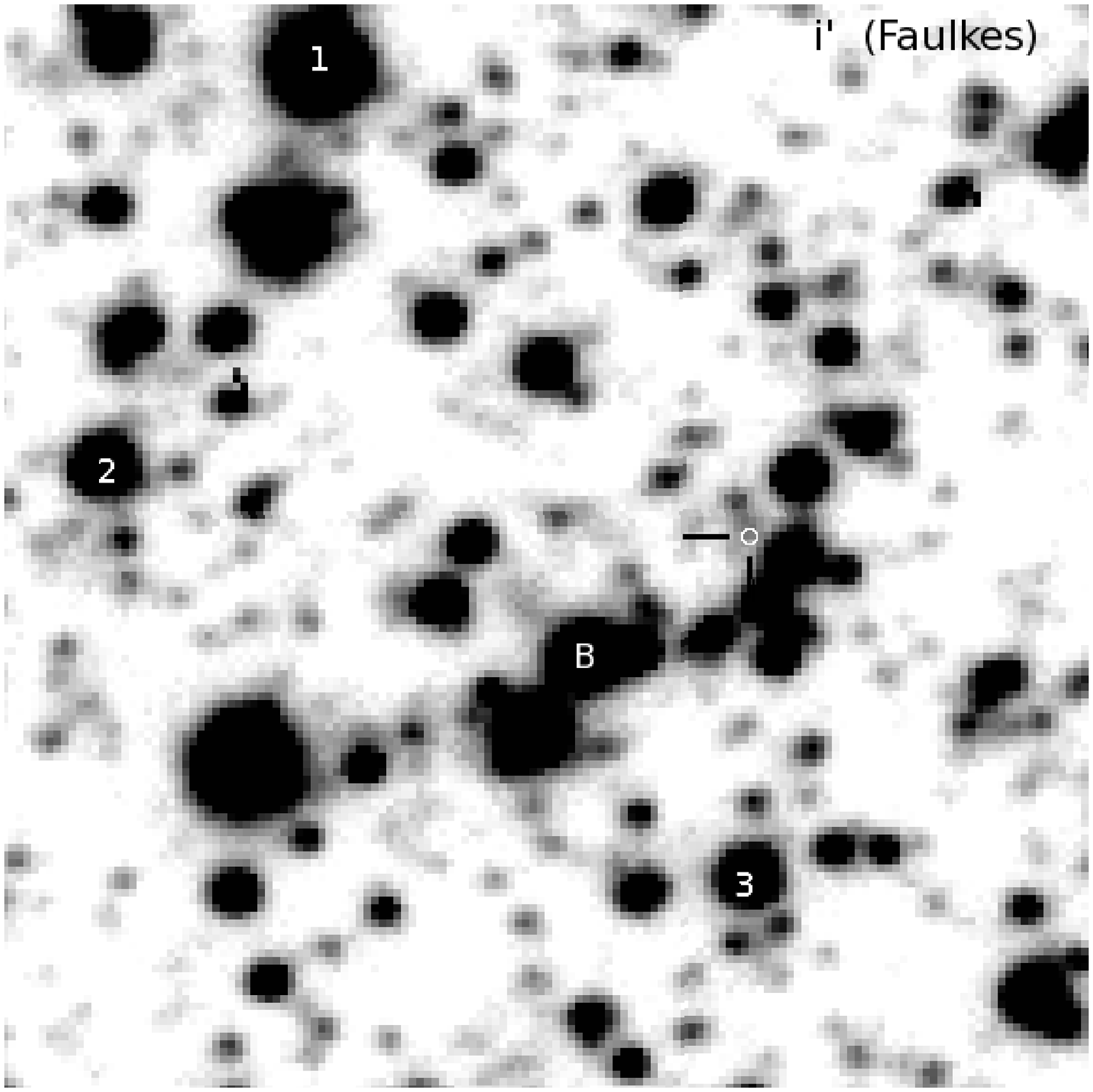}
\includegraphics[width=8.3cm,angle=0]{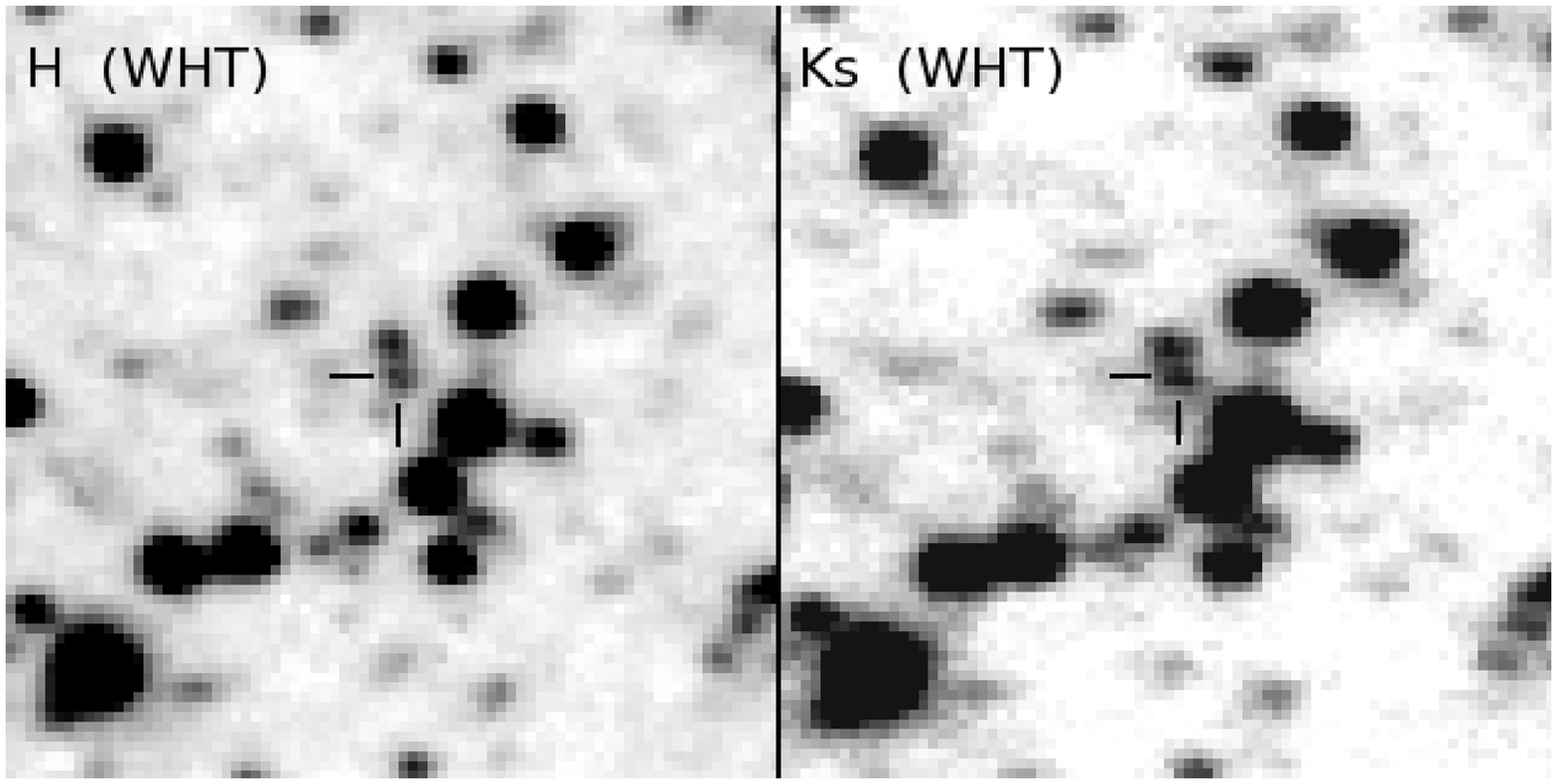}
\caption{High resolution optical and NIR finding charts. North is up, east is to the left. \emph{Upper:} $40\arcsec \times 40\arcsec$ Faulkes i$^{\prime}$-band image from 2010 July 15 (the seeing was 0.9\arcsec). The 0.31\arcsec\ \textit{Swift} UVOT error circle of \1752 (white circle) and star B, are marked (C11), as are three field stars used for relative photometry. A faint field star (R = $19.9 \pm 0.1$; i$^{\prime}$ = $19.2 \pm 0.1$) lies 0.8\arcsec\ to the north of \1752. \emph{Lower:} $20\arcsec \times 20\arcsec$ WHT/LIRIS images on 2010 July 1 in H-band and K$_{\rm S}$-band (seeing 0.6--0.8\arcsec).}
\end{figure}

\section{Observations}

\subsection{Faulkes Telescopes optical monitoring}

The outburst decay of \1752 was regularly monitored with the 2-m Faulkes Telescopes (FTs) North and South, located at Haleakala on Maui and Siding Spring, Australia, respectively, as part of an ongoing monitoring campaign of $\sim 30$ low-mass X-ray binaries \citep{lewiet08}. 
Images in Bessell B, V, R and Sloan Digital Sky Survey (SDSS) i$^{\prime}$-bands (mostly 100-sec exposures each) were made every $\sim 3$ days from 2010 March 22 until 2010 July 15, continuing 
less regularly until 2010 October 20, totalling 206 images in seven months. The cameras EM01 (on FT North) and EM03 (on FT South) were used; both have 2048 $\times$ 2048 pixels binned 2 $\times$ 2 into effectively 1024 $\times$ 1024 pixels. The field of view (FOV) is $4.7\arcmin \times 4.7 \arcmin$ and the pixel scale is 0.278\arcsec pixel$^{-1}$. Bias subtraction and flat-fielding were performed via the automatic pipelines. We identify the optical counterpart of \1752 as a star fading by $\geq 2.6$ magnitudes over the observing period, at a position consistent with that reported by C11; \cite{millet11}. The counterpart is detected until 2010 July 15, after which detections become ambiguous due to its faintness and proximity to field stars of similar magnitude. In Fig. 1 we present high resolution optical and NIR finding charts \citep[for wider field, lower resolution optical finding charts see C11;][]{millet11}.

Photometry was performed on the BHXB and the three field stars using \small PHOT \normalsize in \small IRAF\normalsize, adopting a 0.83\arcsec aperture, optimized to contain the maximum flux from \1752 while minimizing the contamination from the close field stars. For this reason we discarded all data with poor seeing ($> 1.7\arcsec$). We calibrated the field using Landolt standard stars observed on a number of dates and adopted the transformation to SDSS i$^{\prime}$-band from R and I described in \cite*{jordet06}. The error on the absolute calibration is $\sim 0.1$ mag for each filter.
The light curves are presented in Fig. 2 (upper panel) and discussed in Section 3.

\subsection{\textit{Swift} UVOT optical monitoring}

\textit{Swift} UltraViolet/Optical Telescope \citep[UVOT;][]{romiet05} data have been pre-processed at the \textit{Swift} Data Center \citep[see][]{breeet10} and require only minimum user processing. The image data of each filter, from each observation sequence, i.e., with a given observation ID, were summed using {\tt uvotimsum}. Photometry of the source in individual sequences is derived via {\tt uvotmaghist}, using an extraction region of radius 2.5\arcsec\ (to minimize contamination from close field stars; see Section 3). Magnitudes are based on the UVOT photometric system, which differs from the Bessell system by V$-v < 0.04$ and B$-b < 0.04$ for all reasonable spectral indices \citep{poolet08}. $v$ and $b$-band light curves are shown in Fig. 2.

\begin{table*}
\begin{center}
\caption{VLT/VISIR observation log and flux upper limits.}
\begin{tabular}{lllllllll}
\hline
Date&MJD&Airmass&\multicolumn{3}{l}{Total exposure time on source / seconds}&\multicolumn{3}{l}{Derived flux density (mJy; $3\sigma$ upper limits)}\\
    &   &       &8.59 $\mu$m&10.49 $\mu$m&11.96 $\mu$m&8.59 $\mu$m&10.49 $\mu$m&11.96 $\mu$m\\
\hline
2010-07-10&55387.1& 1.02--1.06 & 541 & 707 & 690  &$<$0.96&$<$2.12&$<$1.33\\
2010-07-11&55388.1& 1.02--1.08 & 541 & 707 & 690  &$<$1.11&$<$1.76&$<$1.87\\
2010-07-12&55389.2& 1.12--1.24 & 541 & 707 & 690  &$<$0.71&$<$1.75&$<$1.88\\
Combined data&55387--9& 1.02--1.24 &1623&2121&2070&$<$0.46&$<$1.10&$<$1.04\\
\hline
\end{tabular}
\end{center}
\end{table*}

\begin{figure*}
\centering
\includegraphics[width=14cm,angle=270]{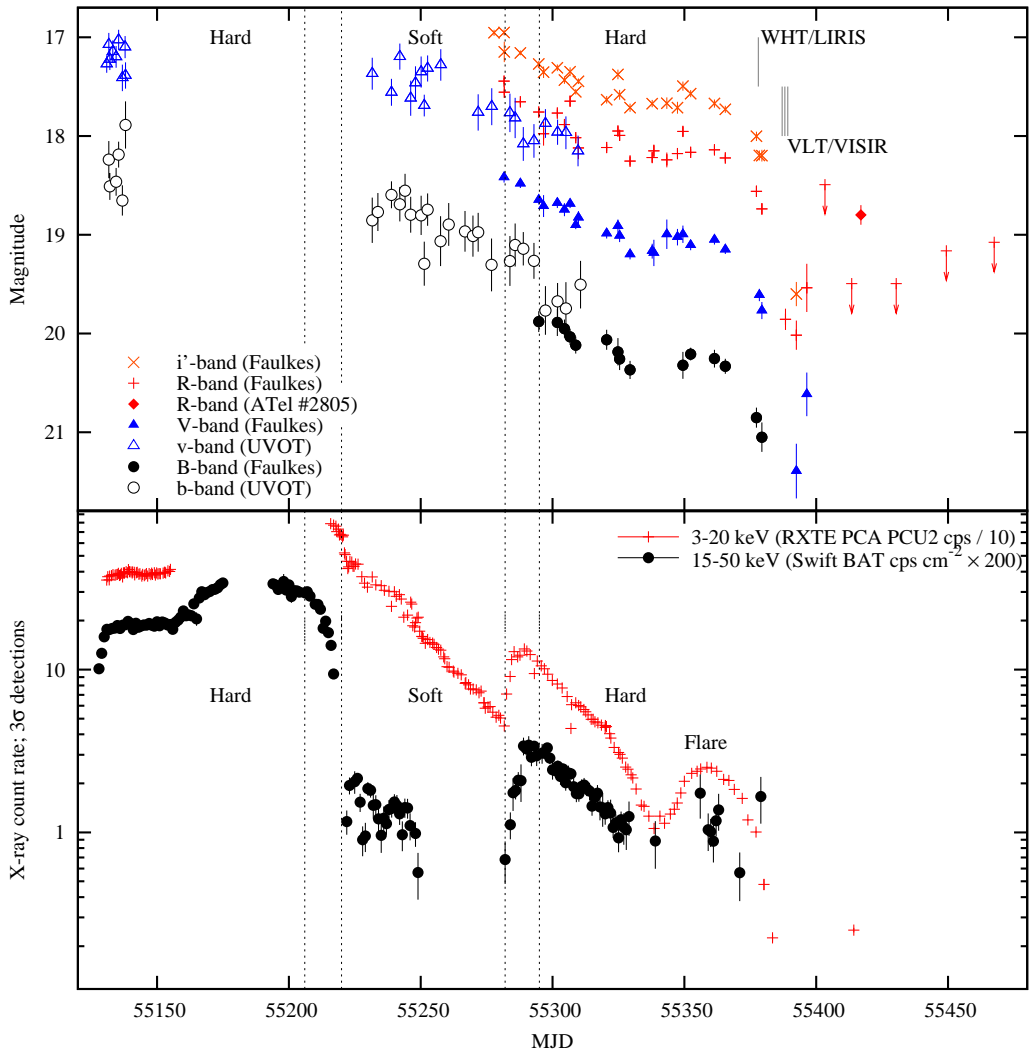}
\caption{Outburst light curves. The dotted vertical lines are X-ray state transitions as reported by \citeauthor{shapet10} (2010). \emph{Upper:} Optical, NIR and mid-IR data. The dates of the mid-IR and NIR observations are marked by vertical black lines. ATel \#2805 refers to \citeauthor{corret10} (2010). \emph{Lower:} X-ray data (3$\sigma$ detections). cps = counts per second.}
\end{figure*}

\subsection{William Herschel Telescope near-infrared data}

On 2010 July 1 (MJD 55378.0) we observed \1752 with the Long-slit Intermediate Resolution Infrared Spectrograph (LIRIS) in imaging polarimetry mode on the 4.2-m William Herschel Telescope (WHT) at the Observatorio del Roque de los Muchachos, La Palma, Spain. Conditions were excellent with a seeing of 0.6--0.8\arcsec. The total on source exposure times were 240 sec in near-IR (NIR) H-band and 360 sec in K$_{\rm S}$-band. 
The Wollaston prism splits the incoming light into four simultaneous images, one at each of the four polarization angles; 0$^{\circ}$, 45$^{\circ}$, 90$^{\circ}$ and 135$^{\circ}$; the FOV for each angle is 4' $\times$ 1'. 
Images were reduced with \small THELI \normalsize \citep{erbeet05} using a series of dome flat fields taken at two dome rotations 90$^{\circ}$ apart, to minimize polarization introduced from the flat fields (biases are determined automatically before every exposure and immediately subtracted). Reduced images 
were aligned and combined in \small IRAF\normalsize. Both the BHXB and the close star 0.8\arcsec\ to the north were detected and resolved (see Fig. 1, lower panels). Total fluxes and polarization Stokes parameters were derived \citep*[see][]{barret08} for \1752 and several field stars in the 2MASS catalogue. 

Using 2MASS stars for flux calibration, we find H = $15.6 \pm 0.1$; K$_{\rm S}$ = $15.2 \pm 0.1$ mag for \1752 (dates are indicated in the light curve; Fig. 2) and H = $15.5 \pm 0.1$; K$_{\rm S}$ = $15.2 \pm 0.1$ for the close star. The level of polarization is $P_{\rm H} < 30$ per cent in H-band and $P_{\rm Ks} < 29$ per cent in K$_{\rm S}$-band (3$\sigma$ upper limits). Instrumental polarization is measured to be no greater than $\sim 1$ per cent \citep{barret08}. These results are discussed in Section 3.7.

\subsection{Very Large Telescope mid-infrared data}

Mid-IR observations with the Very Large Telescope (VLT) were acquired on 2010 July 10, 11 and 12 (Programme ID 086.D-0610). The VLT Imager and Spectrometer for mid Infrared (VISIR) instrument on UT3 (Melipal) was used in small-field imaging mode, with a FOV of $19.2\arcsec \times 19.2\arcsec$ and a pixel scale of 0.075\arcsec\ pixel$^{-1}$. Observations were performed in three mid-IR filters; PAH1 (8.2 -- 9.0 $\mu$m), SIV (10.3 -- 10.7 $\mu$m) and J12.2 (11.7 -- 12.2 $\mu$m). The observation log and results are shown in Table 1. With overheads, the total time including chopping/nodding was 65 min on each date. The data were reduced using the VISIR pipeline. Raw images were recombined and sensitivities were estimated based on standard star observations (the calibrators HD133774, HD178345 and HD217902 were observed).
The rms we measure in the images agrees (typically to within 10--30 per cent) with the expected sensitivities estimated from the standard stars. \1752 was not detected; the deep upper limits given in Table 1 are derived from the rms in a region centred on the position of \1752.

\subsection{X-ray data}

The RXTE Proportional Counter Array (PCA) and \textit{Swift} Burst Alert Telescope (BAT) both monitored \1752 during its outburst. The \textit{Swift} BAT 15--50 keV count rates from the hard X-ray transient monitor are made publicly available by the \textit{Swift} BAT team. The RXTE PCA 3--20 keV X-ray net count rate was extracted with standard procedures \citep*[see e.g.][]{munoet10,mottet10,stieet11}. The count rate light curves (3$\sigma$ detections) are shown in Fig. 2 (lower panel).

In order to perform spectral analysis in the decaying hard state, we fitted PCA spectra between 3 and 40 keV. \textsc{xspec} V. 11.3.2 was used to fit the spectra. We used the \emph{phabs model} in \textsc{xspec} to describe absorption. Following  \cite{munoet10} we used an absorbed (equivalent Hydrogen column value of $n_\mathrm{H} = 7.2 \times 10^{21}$ cm$^{-2}$) broken power law plus a gaussian component to describe a broad iron emission line\footnote{Using a slightly different absorption of $n_{\rm H} = 5.13 \times 10^{21}$ cm$^{-2}$ (C11) makes a negligible difference to the resulting fit parameters.}.  The centroid energy of the line was restricted to 6.4 and 6.7 keV and the line width was restricted to 0.01 and 0.5 keV to prevent artificial broadening due to the response of XTE/PCA around 6.4 keV. No additional reflection component is required to obtain a good fit and no high energy cutoff below 40 keV was needed. Where it was not possible to constrain the energy of the break of the broken power law accurately, we used a simple power law to fit the spectra. Only in some cases the iron line was not necessary to obtain acceptable fits.

Following the method of \cite{bellha90}, the RXTE PCA fractional rms was computed within the frequency bands 0.1--64 Hz and 0.005--1 Hz. The latter, low frequency band was used to achieve accurate rms estimates even at low PCA count rates. PCA channels 0--35 (2--15 keV) were used for computing the rms.

\section{Results}

\subsection{The outburst decay}

The outburst was initially monitored at optical and UV frequencies with \textit{Swift} UVOT (C11). Near the start of the outburst \1752 was bright; V $\sim 17$ (at the end of 2009 October). The Faulkes Telescope optical monitoring began on 2010 March 22 (MJD 55277) when the BHXB was in a soft X-ray state; the source had faded to V $\sim 18.4$ by this time (both UVOT and FT light curves are shown in Fig. 2). The optical outburst morphology can be described by a gradual exponential decay (although there is a gap in the coverage) followed by a plateau, then a faster fade towards quiescence with evidence for reflares. This outburst morphology is similar to outbursts of some other BHXBs. For example, optical exponential decays followed by a plateau and a rapid fade were seen in A0620--00 and XTE J1550--564 \citep{kuul98,jainet01}, and prominent reflares were seen in GRO J0422+32 and GRS 1009--45 \citep{callet95,chenet97}.

\1752 made a transition to the intermediate state and finally to the hard state during the period 2010 March 27 -- April 4 \citep[MJD 55282 -- 55295;][]{shapet10}. The transition is also visible by an increasing X-ray flux, especially in the \textit{Swift} BAT hard X-ray light curve (Fig. 2, lower panel). In most BHXB outbursts this increase of hard X-ray flux is not significant below energies of $\sim 10$ keV \citep[e.g.][]{dunnet10}. In a few cases, like here for \1752 \citep[and e.g. the outbursts of XTE J1650--500 and XTE J1817--330;][]{dunnet10}, the soft X-ray flux also increases over the soft to hard transition \citep[see also][]{shapet10,stieet11}, presumably due to a relatively brighter power law component compared to the disc in these outbursts.

An OIR brightening and reddening is seen after the soft-to-hard transition in some BHXBs, indicative of an increasing synchrotron jet contribution \citep[e.g.][]{buxtba04,coriet09,russet10a}. 
No increase in flux is seen in our optical light curves either during the transition (which implies the X-ray power law component does not extrapolate and contribute much to the optical flux) or after the transition (as we may expect for a jet). Instead, some time after the transition to the hard state is complete there is a plateau, where the optical flux remained at approximately a constant level. In Section 3.3 we assess whether this plateau could be related to the jet. A NIR brightening was reported around the same time of the state transition itself, which may be due to the jet \citep{kaleet11}. On the hard state rise, the jet may have made a strong contribution in the optical (C11).

The optical counterpart continued to fade slowly until around 2010 May 13 (MJD 55329), dropping by $\sim 0.8$ mag in two months, with evidence for variability on day-timescales (changes up to $\sim 0.4$ mag in a few days). \1752 remained at an approximately constant flux level (the plateau) for the next 1.5 months until $\sim$ 2010 June 27, then faded rapidly by $\sim 1.5$ -- 2 mag over the next 14--18 days to R $= 19.9 \pm 0.1$ by 2010 July 11 (MJD 55388) and R $= 20.0 \pm 0.1$; V $= 21.4 \pm 0.3$ by 2010 July 15 (MJD 55392). The NIR detections with WHT/LIRIS, and the mid-IR non-detections with VLT/VISIR, were made just before and at the end of the rapid optical fade, respectively (Fig. 2). An episode of brightening in 2010 August (also plotted in Fig. 2) was reported by \cite{corret10}.

The state transitions are clear in the X-ray light curve \citep[see also C11;][]{shapet10,stieet11} as a prominent fade and subsequent brightening of the \textit{Swift} BAT hard X-ray flux. Near the end of the outburst a reflare is seen in the RXTE PCA light curve lasting $\sim 30$--40 d, just before both optical and X-ray rapidly dropped towards quiescence.

For 28 d over the soft-to-hard transition both UVOT and FT were monitoring the optical counterpart. While both light curves show a slight fading during this time, there is a difference of $\sim 0.7$ mag between FT V-band and UVOT $v$-band on the same dates, which is too large to be accounted for by differences between the Bessell and UVOT photometric systems \citep{poolet08}.  There are at least six stars located $< 4$\arcsec\ from \1752 (Fig. 1), which are unresolved in the UVOT images (C11), so it is very likely that these stars contaminated the UVOT light curve when the source faded, even though we used a 2.5\arcsec\ extraction radius to minimise such contamination. Contamination from nearby stars within the larger UVOT aperture probably accounts for the discrepancy. Comparing the FT and UVOT magnitudes of star B we find that they differ by only 0.01 mag in B ($b$) and 0.04 mag in V ($v$) when using the same size aperture (4\arcsec), so the discrepancy cannot be due to a systematic calibration issue.

\subsection{Spectral energy distributions}

In order to assess the disc and jet contributions to the OIR \citep[C11;][]{kaleet11} and mid-IR fluxes, we construct SEDs from various stages of the outburst (Fig. 3). To derive the de-reddened fluxes we adopt an interstellar absorption of $n_{\rm H} = (5.13 \pm 0.03) \times 10^{21}$ cm$^{-2}$ (C11) which corresponds to an optical extinction \citep{predet95} of $A_{\rm V} \sim 2.87$ \citep*[we use the extinction law of][]{cardet89}.

Radio, NIR and optical observations were taken within five days of each other, soon after the discovery of the outburst, when \1752 was in a bright, hard state \citep[][C11]{brocet09,torret09b,millet11}. We plot these data in Fig. 3 (SED [1]; black crosses). The OIR emission is consistent with a power law of spectral index $\alpha_{\rm OIR} \sim +1.0$ from B-band to K-band, a blue spectrum indicative of a blackbody, very likely the outer regions of the (possibly irradiated) accretion disc \citep[e.g.][]{hyne05}. The extrapolation of the flat spectrum seen at radio frequencies ($\alpha_{\rm radio} \sim 0$) is close to the K-band flux density, so the jet is unlikely to contribute much of the emission above this frequency. An optical SED from one date soon after, in the hard state \citep[using data from C11;][]{millet11} appears slightly bluer (SED [2]; purple open squares). An optical SED from the soft state (FT V,R,i$^{\prime}$-bands) is also blue; $\alpha_{\rm OIR} \sim 0.5$ (SED [3]; orange diamonds in Fig. 3).

\begin{figure}
\centering
\includegraphics[width=5.8cm,angle=270]{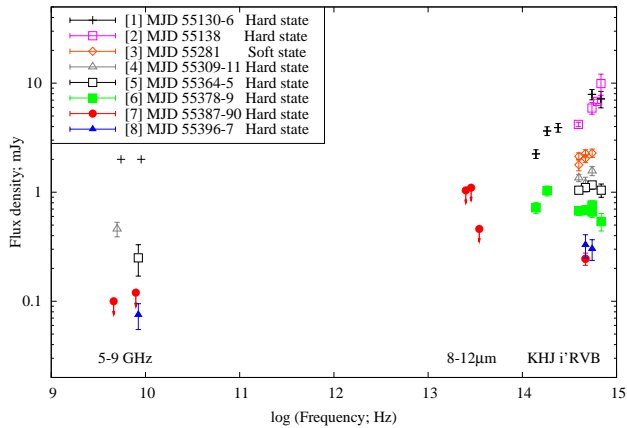}
\caption{Broadband, unabsorbed radio to optical SEDs at various stages of the outburst (see text for details). The radio data ($\nu < 10^{10}$ Hz) were taken with the ATCA \citep[black crosses;][]{brocet09}, VLBA \citep[grey open triangles and black open squares;][]{millet11} and EVLA (red solid circles and blue solid triangles; Jonker et al. in preparation).}
\end{figure}

During the hard state outburst decay, radio data were acquired \citep{millet11} and two detections were within a few days of FT optical observations. These SEDs are shown as grey open triangles and black open squares in Fig. 3 ([4] and [5]). The observed radio to optical (i$^{\prime}$-band) spectral index for both is $\alpha = +0.1$, slightly inverted, and typical of hard state BHXBs \citep{fend01,russet06}.

The NIR WHT data were acquired within 1.5 days of FT B,V,R,i$^{\prime}$ data during the hard state decay; this SED ([6]; green solid squares in Fig. 3) is fainter and redder ($\alpha_{\rm OIR} \sim -0.2$ between B and K$_{\rm S}$-bands) than the SED on the hard state rise, as may be expected as the outer disc cools and reddens. However, the OIR SED is too flat for an outer disc, and a peculiar excess in H-band is hard to account for. The SED may originate in the jet \citep[as also suggested by][from NIR data taken during the soft-to-hard transition]{kaleet11} or perhaps partly from the companion star. If the jet is dominating the NIR excess, the H-band flux is too bright for a standard flat or optically thin synchrotron spectrum. However, similar SEDs have been reported \citep[e.g.][]{lewiet10} and the spectral shape of the NIR excess could be explained by the jet base \citep*{market05} or a cooling feature in the jet \citep{peerca09}. SED [1] also shows a similar H--K$_{\rm S}$ colour, but that SED was acquired from a different telescope so it is unlikely to be a common calibration issue (and the effect of an incorrect extinction cannot account for the spectral shape here in the IR). The exposure times of the non-simultaneous H and K$_{\rm S}$ observations were a few minutes, so short-term variability may dominate the fluxes \citep[OIR variability can be high on these timescales;][]{caseet10,gandet10} and confuse this SED. If the companion dominates, the K$_{\rm S}$, V and B absolute magnitudes \citep[at $\sim 3.5$ kpc;][]{shapet10} and H--K$_{\rm S}$ colour are consistent with an F star \citep{ruel91}. Since the star is suggested to be later than G5V (Section 3.8), and the optical flux continued to fade, the star probably did not dominate the NIR here. This SED is revisited in Sections 3.3 and 3.6.

In a paper in preparation, Jonker et al. report radio monitoring with the EVLA around the time of the VLT/VISIR observations: $< 100$ $\mu$Jy at 4.6 GHz and $< 120$ $\mu$Jy at 7.9 GHz on 2010 July 13 (MJD 55390.2; 3$\sigma$ upper limits), and $75 \pm 20$ $\mu$Jy at 8.4 GHz on 2010 July 20 (MJD 55397.2). These are shown in SEDs [7] and [8] in Fig. 3. The array was in its most compact `D' configuration for all observations, and the angular resolution was $\sim 7.5\arcsec$ at 8.4 GHz, which is insufficient to distinguish the core jet from downstream shocks.  The whole bandwidth at a single frequency was required to detect \1752, so spectral constraints were not possible. Downstream jet shocks (rather than a core jet) like those seen at earlier times in the outburst \citep{yanget10,millet11} therefore cannot be empirically ruled out. However, the deceleration of the jet components was already observed up until MJD 55253 \citep{yanget10} and these components continued to fade, and the core jet returned and brightened over the soft to hard transition around MJD 55311, becoming brighter than the previously launched ejecta \citep{yanget11}. Further external shocks are unlikely since the bulk of the energy had already been dissipated in the first set of decelerating shocks, so it is likely that the radio emission seen by the EVLA is dominated by the reignited core jet.

An optical R-band detection with FT was made 0.2 d after the VLT/VISIR observation on 2010 July 11. Although the latter radio detection was nine days after our mid-IR observations, FT V,R,i$^{\prime}$ observations within 24 hours of the radio detection confirm the source was at a similar optical flux level on these two dates (SEDs [7] and [8]; this is after the rapid optical fade between MJD 55379 and 55388). In the hard state, the radio and optical luminosities are known to be correlated and powered by the mass accretion rate \citep[e.g.][]{fend01,gallet03,homaet05,russet06}. Therefore, the radio flux on the date of the VISIR observations is likely to be similar to that measured on 2010 July 20. The radio detection and our mid-IR upper limits infer a jet spectral index from radio to mid-IR of $\alpha_{\rm radio-midIR} < 0.25$. The radio to optical (R-band) spectral index is $\alpha_{\rm radio-optical} = 0.11 \pm 0.04$.

Broadband spectra of BHXB jets at low luminosities have only been constrained in a few systems, so these measurements provide a useful addition \citep[the radio to mid-IR 8$\mu m$ spectral index of A0620--00 in quiescence is $\alpha \sim 0.22$;][]{gallet07}. From our SEDs of \1752 we can conclude that either the optically thick jet spectrum is not highly inverted ($\alpha < 0.25$) or the jet break between optically thick and optically thin synchrotron lies at longer wavelengths than 8.6$\mu m$. At a putative distance of $3.5 \pm 0.4$ kpc  \citep{shapet10}, the radio detection would correspond to an X-ray luminosity of $\sim 2 \times 10^{32}$ erg s$^{-1}$ (2--10 keV) if it were to follow the X-ray--radio correlation of hard state BHXBs \citep{gallet03}. This would correspond to a possible bolometric luminosity \citep[see][]{miglfe06} of $\sim 10^{-6} L_{\rm Edd}$ at this time \citep[adopting a BH mass of $\sim 10 M_{\odot}$;][]{shapet10}. However, it is worth noting that \1752 is a radio-quiet BHXB \citep[a number of radio-quiet BHXBs have now been reported; e.g.][]{calvet10,coriet11,solefe11}. The 2 mJy radio detection in the initial hard state \citep{brocet09} on MJD 55136 is $\sim$ one order of magnitude too faint at this X-ray flux (which we take from Fig. 2) to lie on the canonical radio--X-ray correlation of BHXBs.

\begin{figure}
\centering
\includegraphics[width=8.2cm,angle=0]{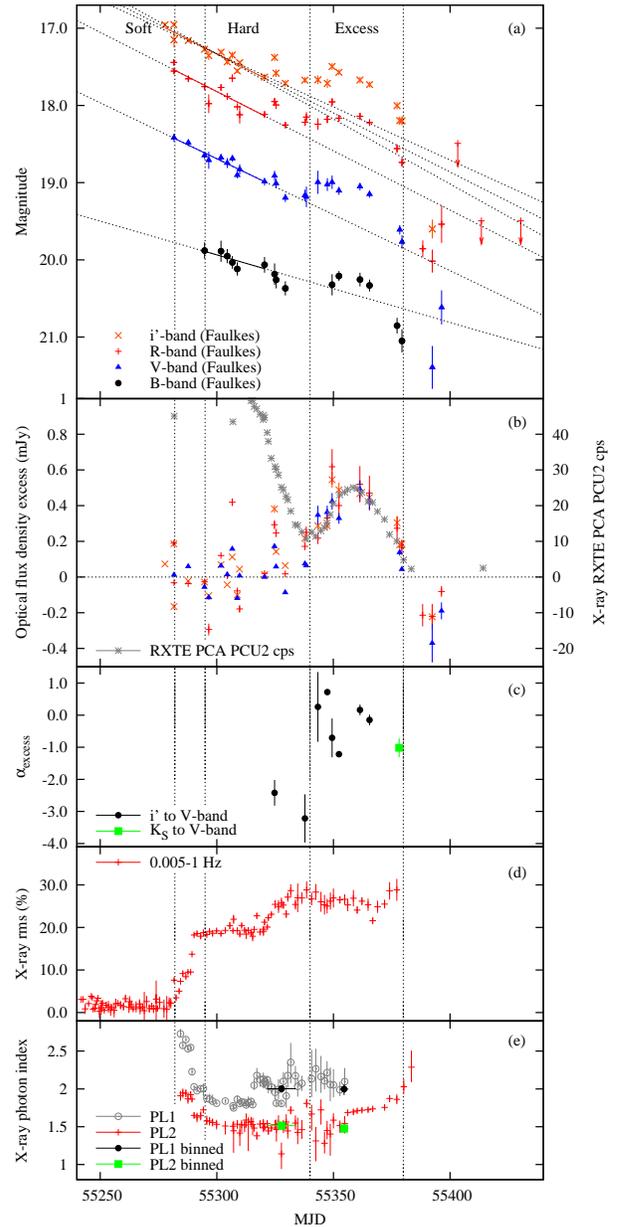}
\caption{\emph{(a)} Exponential decays fitted to the optical (FT) light curves up until MJD 55321. The errors on the slope of the fit are shown for i$^{\prime}$-band. \emph{(b)} The residuals (data minus fit) assuming an extrapolation of these fits to later times, and the X-ray RXTE light curve overplotted. Errors on the residuals are propagated from the fit slope errors and the observed flux errors. The morphology of the observed X-ray light curve is remarkably similar to that of the optical excess light curve (see text). \emph{(c)} Spectral index $\alpha$ of the optical excess, estimated by fitting the individual V,R,i$^{\prime}$ SEDs. \emph{(d)} The X-ray rms variability in the frequency range 0.005--1 Hz. \emph{(e)} The estimated X-ray power law photon index $\Gamma = 1 - \alpha$. For most observations a broken power law is fitted; both photon indices are shown. The vertical lines in all panels indicate the soft to hard state transition and the X-ray flare associated with the optical excess.}
\end{figure}

\subsection{Separating the disc and jet emission}

An additional method to assess a possible synchrotron jet contribution in the hard state decay of the outburst was presented in \cite{russet10a}. There, in the BHXB XTE J1550--564 the OIR light curve was described by an exponential decay in the soft state, and the extrapolation of this decay was subtracted from the hard state light curve to estimate the excess light, which was found to be from the jet. We see in Fig. 4a that the optical light curves of \1752 may also be approximated by an exponential decay and the aforementioned plateau appears to be an excess above this decay.

We fit the optical FT light curves before MJD 55321 with an exponential decay, and estimate the excess in the hard state decline (Fig. 4a). There are fewer data in B-band and the flux of the excess in that band is hard to estimate due to the uncertainty in the slope of the fit. In V, R and i$^{\prime}$-bands there is scatter of $\sim 0.2$ mJy around the fit before 55321 which appears, in most cases to be uncorrelated between the bands (Fig. 4b). A clear excess above the fit is apparent after MJD $\sim 55340$ in all three filters, peaking at $\sim 0.5$ mJy and lasting $\sim 30$ -- 40 d. Near the end of the outburst, after MJD $\sim 55384$ the data drop below the extrapolation of the fit.

The exponential decay is likely to be due to the fading accretion disc, which has a blue SED, as measured in the soft state (see Section 3.2). The excess may be jet emission, which is expected to make a contribution in the hard state but not in the soft state. The disc component eventually fades faster than the extrapolation of the exponential decay since the data at late times appear to lie below this extrapolation. This may result from a decreasing outer disc temperature and/or a lack of irradiating photons ionizing the disc at these low luminosities. The latter is consistent with the X-ray light curve, which drops rapidly at this time (Fig. 1).

Remarkably, the morphology of the optical light curve of the excess is very similar to the simultaneous X-ray 3--20 keV light curve (shown overplotted in Fig. 4b). The X-ray flare is simultaneous with the optical excess, and it is immediately apparent that the two are correlated (see also Section 3.4), the optical excess and the X-ray count rate both brightening then fading together (Fig. 4b). The spectral index of the optical excess seems to be redder than the disc component (see Section 3.2), and could be variable (Fig. 4c). However, the three optical filters only cover 0.14 orders of magnitude in frequency and any short term variability could dominate the individual SEDs since observations in the three wavebands were not strictly simultaneous. The average spectral index of the excess during the main flare, taking into account errors, is $\alpha = -0.16 \pm 0.29$. This slightly favours a flat spectrum, self-absorbed jet ($\alpha \sim 0.0$, as is expected between radio and IR) rather than an optically thin jet ($\alpha \sim -0.6$).

Just before the main flare, when the excess is significant its spectral index is redder; $\alpha = -3 \pm 1$. The spectral index of the OIR jet in XTE J1550--564 was found to start quite red, $\alpha \sim -1.5$ then smoothly increase until it levelled off at $\alpha \sim -0.6$ in the declining hard state \citep{russet10a}. We find evidence for this to also be the case for \1752 -- the spectral index of the excess increases as it brightens. There is also evidence for a change of spectral index of the excess near the end of the flare. On MJD 55378, the NIR detections (Section 2.3) are around $\sim 1$ mJy (Fig. 3). If the exponentially decaying disc has a spectral index $\alpha \sim +1$ (Section 3.2), we estimate the disc flux would be around $\sim 20$ -- 30 per cent of the observed NIR flux. This fraction will be lower if the disc does indeed fade faster than the exponential decay at these late times. Alternatively, this fraction could be higher if the disc reddens as it fades \citep[e.g.][]{maitba08}. Since irradiation is unlikely to be important (see above), we may expect a spectral index as low as $\alpha = +1/3$ for a standard disc. As a conservative estimate, we impose a limit $\alpha \geq 0.0$ for the disc and obtain an excess (non-disc) NIR flux of $0.75 \pm 0.39$ mJy in H-band and $0.44 \pm 0.37$ mJy in K$_{\rm S}$-band. This yields a power law fit to the optical and NIR SED near the end of the flare of $\alpha = -1.0 \pm 0.3$ (shown in green in Fig. 4c), which appears to be redder than the average optical spectral index during the flare, $\alpha = -0.16 \pm 0.29$. Instead, the spectral index is more consistent with optically thin synchrotron, so the jet break may shift to lower frequencies near the end of the flare, possibly indicative of the jet losing power. This spectral index is also too red and/or too bright to be explained by an F or G-type star, so the companion cannot provide a strong contribution.

The optical flare is unlikely to be due to irradiation on the surface of the disc or the companion star. The mean spectral index of the flare is much redder than the Rayleigh-Jeans tail of the blackbody due to irradiation, for which we expect $\alpha = +2$. The peak of the blackbody due to irradiation would have to lie in the optical or at lower frequencies to explain the data. If this were the case, the temperature of such a blackbody would be $T_{\rm irr} \leq 5300$ K. The optical luminosity (V + R + i$^{\prime}$) of the excess when it is brightest is $(1.08 \pm 0.08) \times 10^{34}$ erg s$^{-1}$ and the X-ray (3--20 keV) luminosity from RXTE PCA at that time is $(3.61 \pm 0.04) \times 10^{35}$ erg s$^{-1}$ \citep[at a distance of 3.5 kpc;][]{shapet10}, so $\geq 3$ per cent of the X-ray luminosity would have to be intercepted by the disc and efficiently reprocessed in the optical regime. This fraction is probably unphysically high; in other BHXBs an irradiation fraction has been measured, and is generally $\sim 0.3$ per cent or less \citep*{gieret09,chiaet10,zuriet11}. Irradiation therefore cannot easily explain the optical flare. The spectral index of the flare, and its appearance in the decaying hard state \citep[similar to other BHXBs, like XTE J1550--564, GX 339--4 and 4U 1543--47;][]{jainet01,buxtba04,coriet09,russet10a} favour a jet origin. In Section 3.5 we assess the likelihood that the jet also produces the simultaneous X-ray flare.

\begin{figure}
\centering
\includegraphics[width=8.2cm,angle=270]{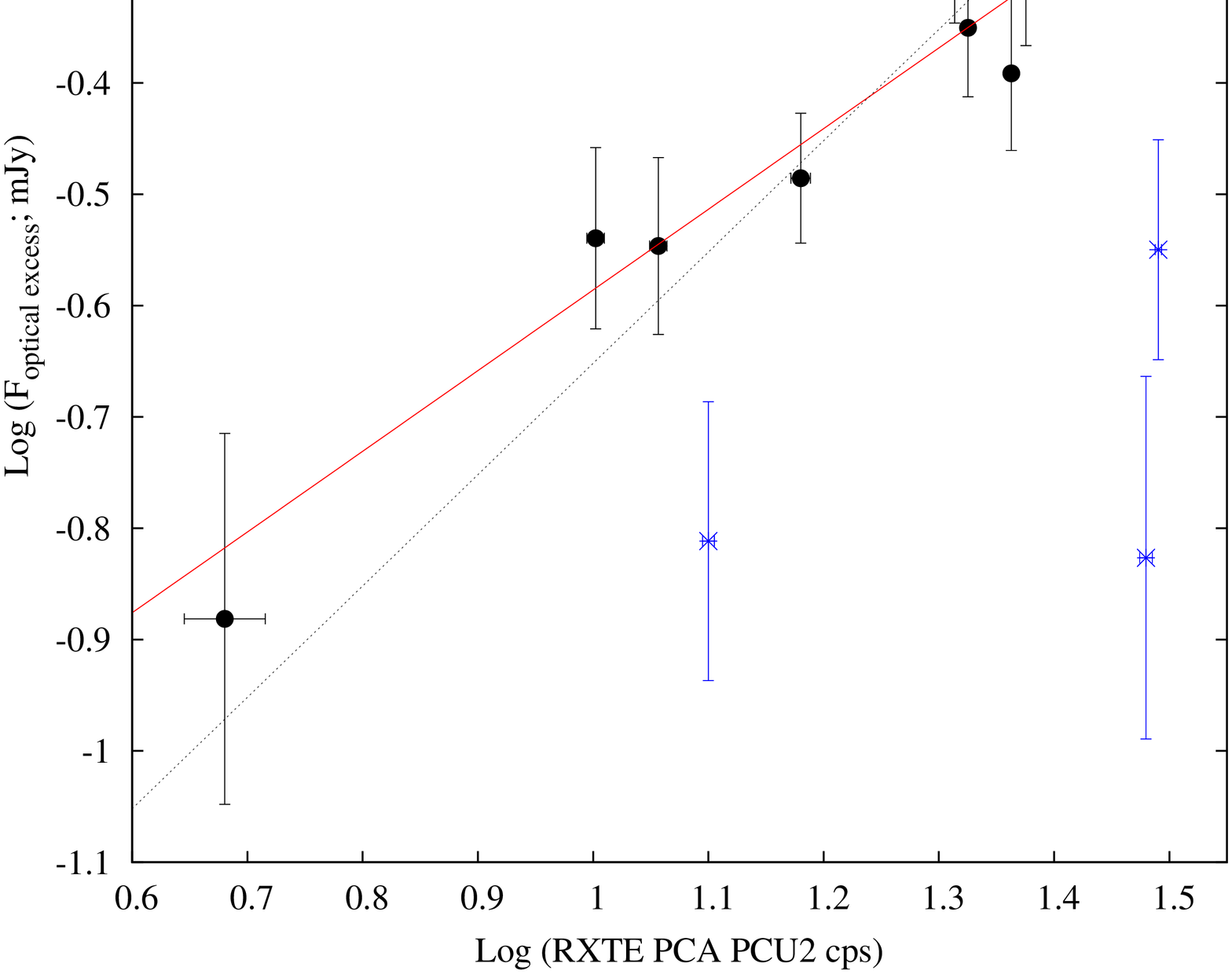}
\caption{X-ray counts (3--20 keV) versus optical excess jet emission (defined as the excess above the optical exponential decay in the light curve). The average flux density of the excess from V, R and i$^{\prime}$-bands is used. The power law fit is to the main flare seen in the optical excess and in the X-ray band, and excludes the data before this flare (before MJD 55340).}
\end{figure}

\begin{figure*}
\centering
\includegraphics[width=6.1cm,angle=270]{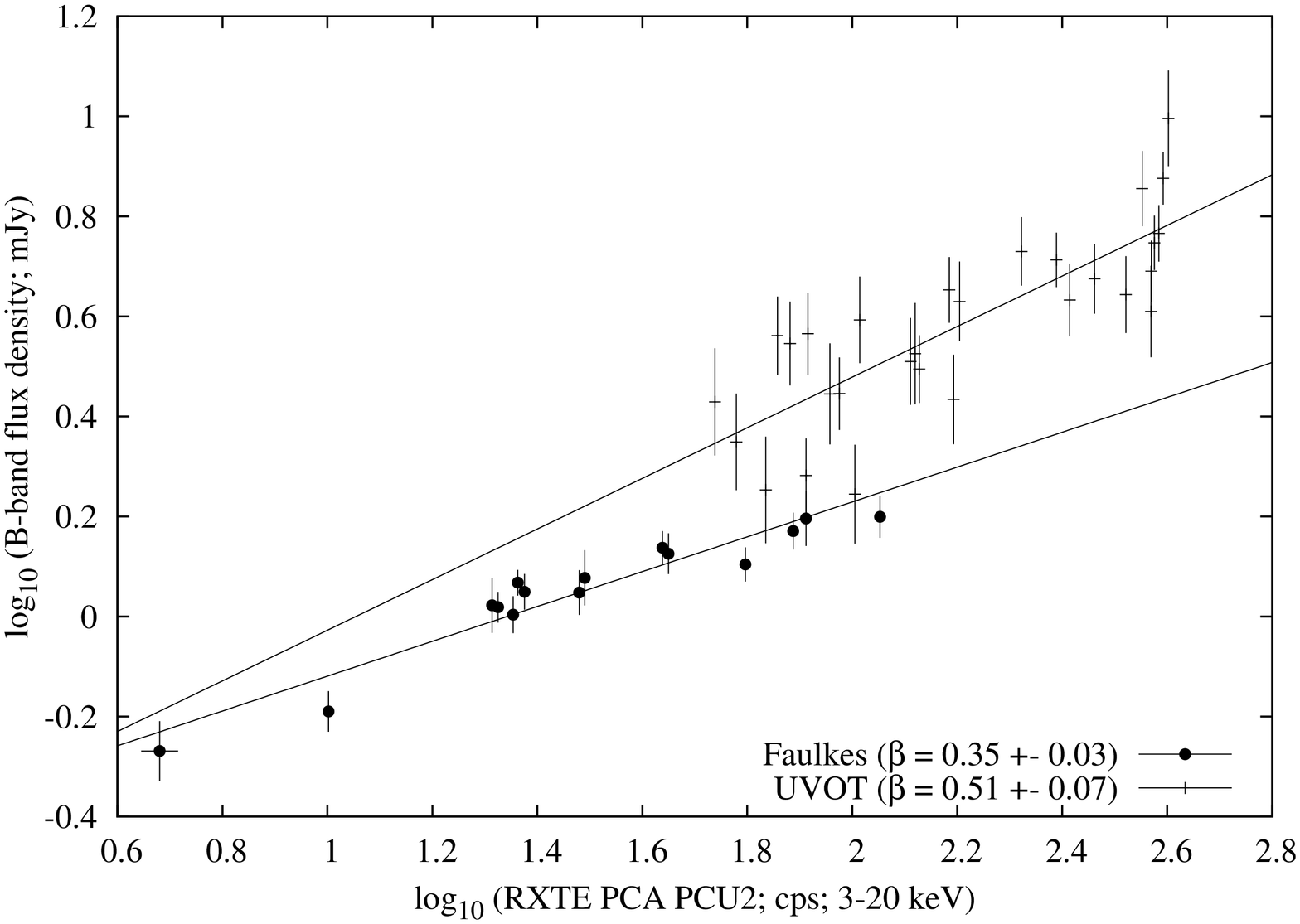}
\includegraphics[width=6.1cm,angle=270]{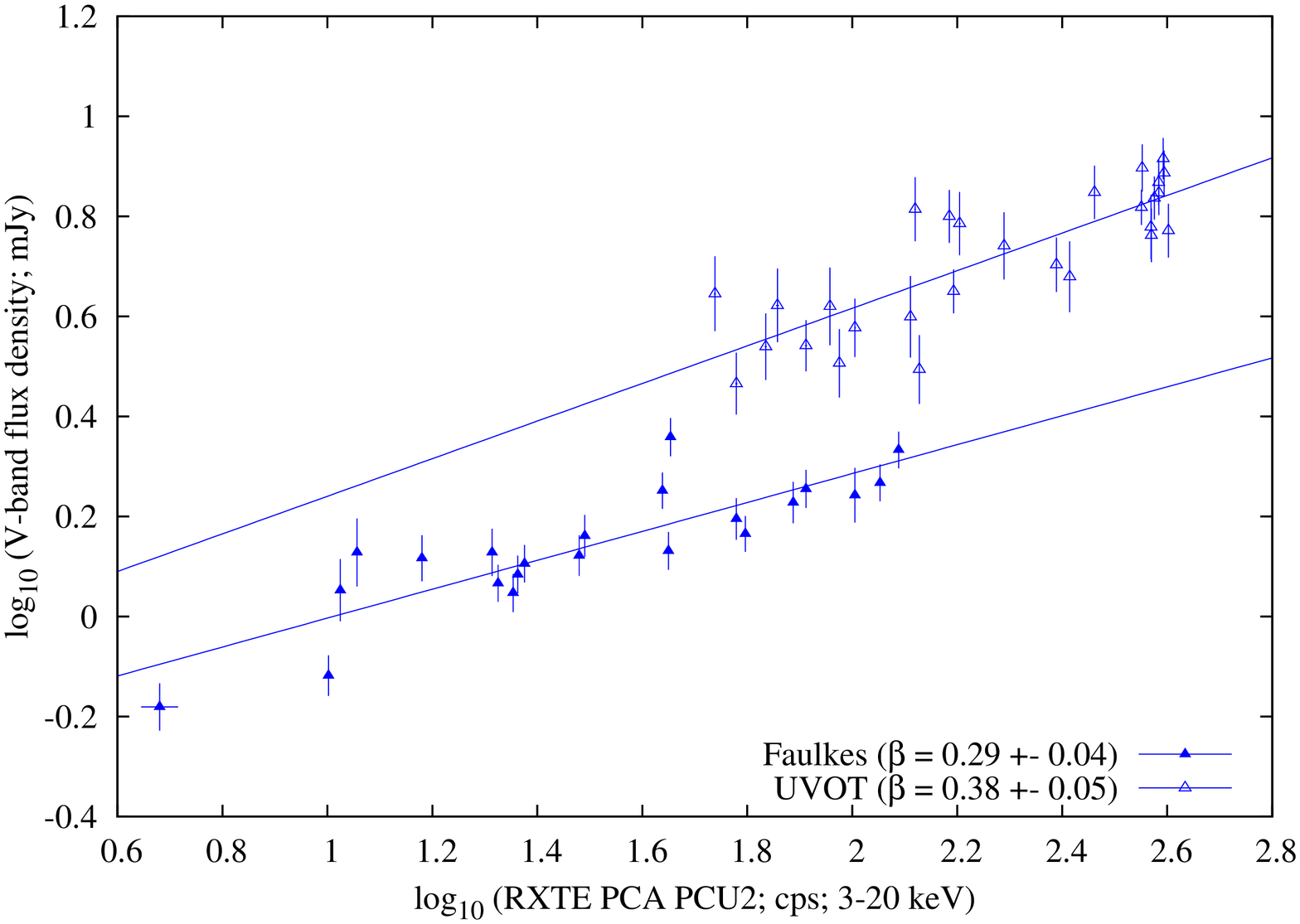}\\
\includegraphics[width=6.1cm,angle=270]{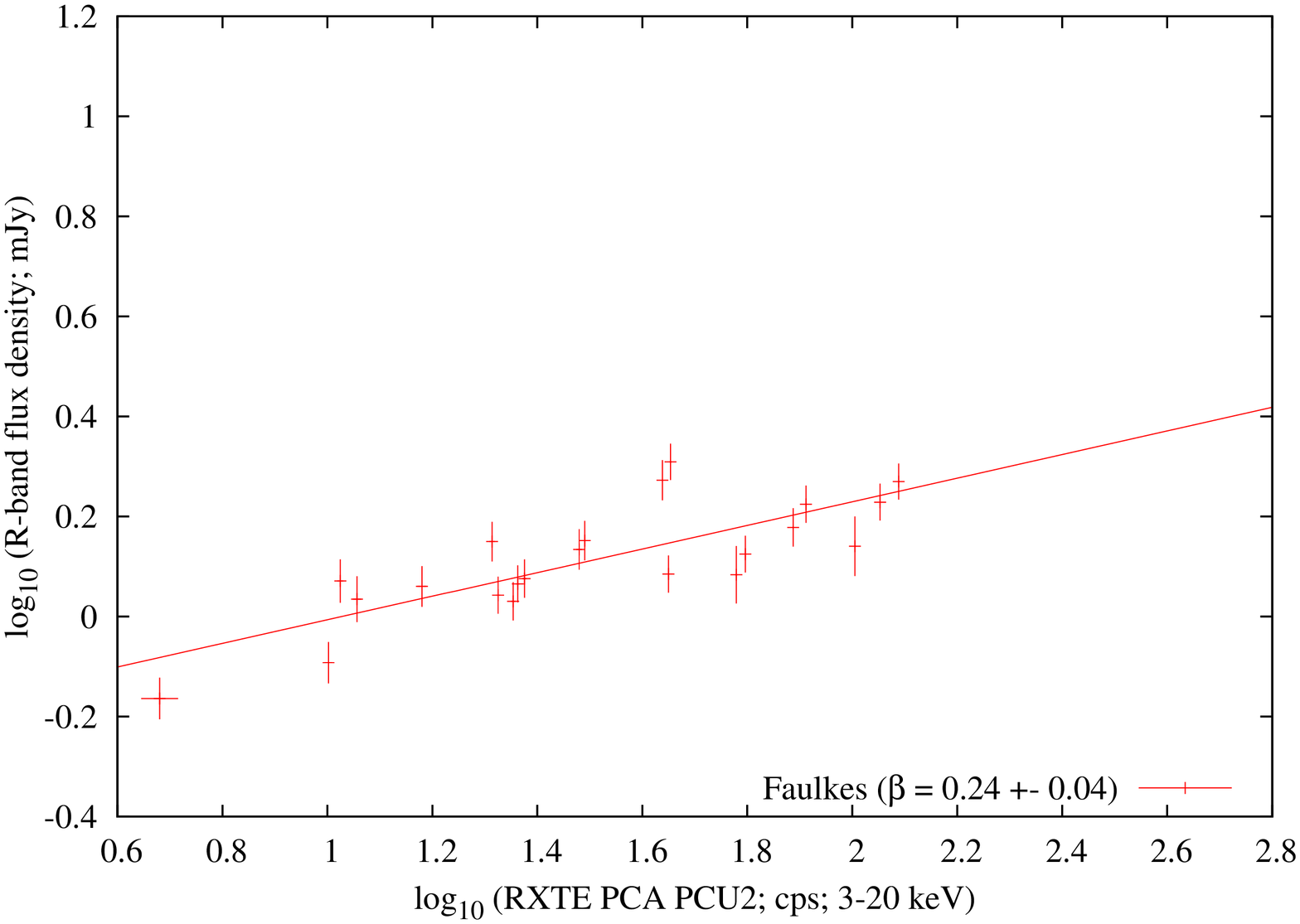}
\includegraphics[width=6.1cm,angle=270]{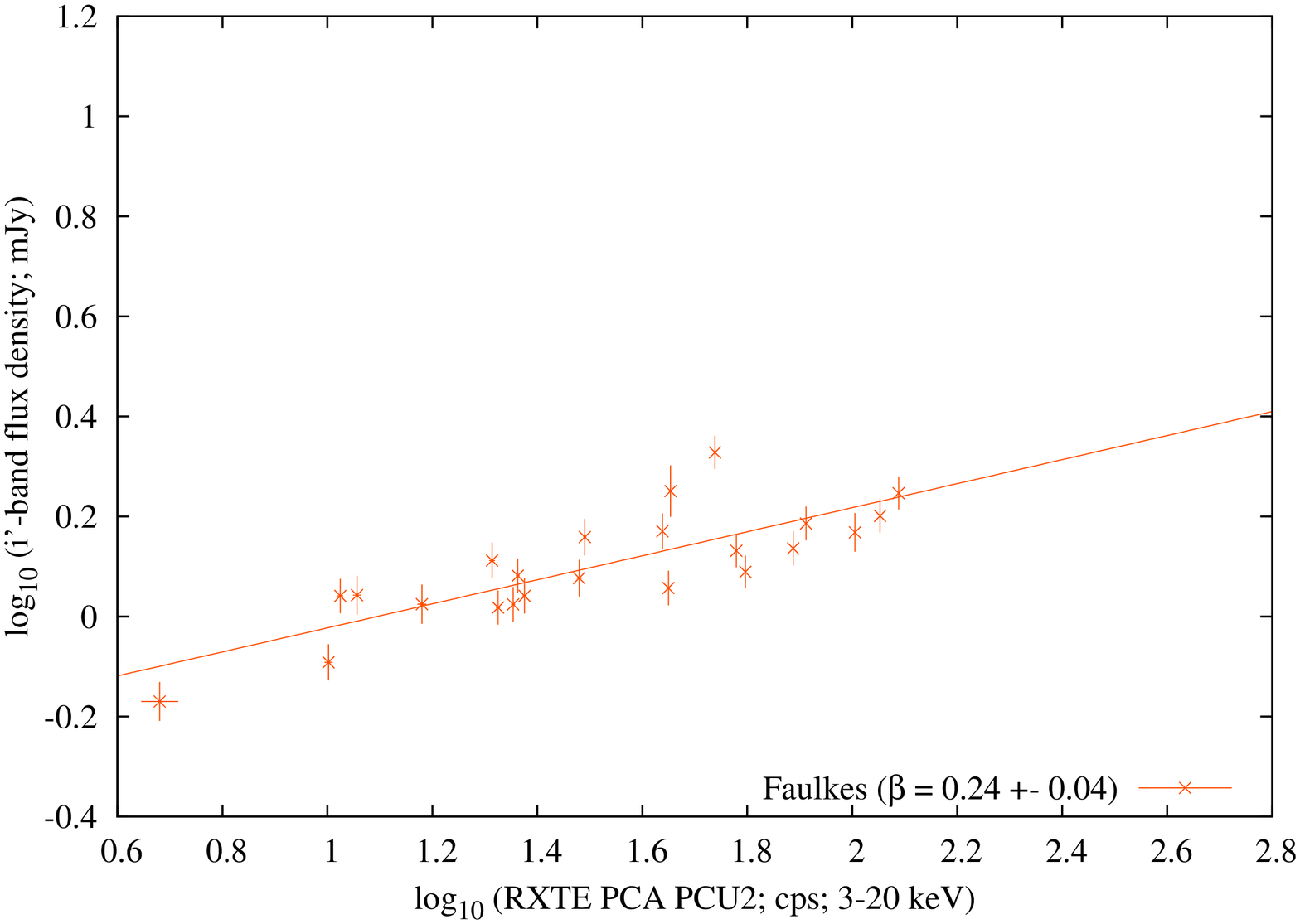}\\
\caption{X-ray versus optical (total, disc plus jet) count rates for the four optical wavebands. For comparison the same scale is used for all plots. The UVOT data are from the first half of the outburst, the Faulkes data are from the second half (there is some overlap).}
\end{figure*}

\subsection{Optical--X-ray correlations}

It was shown that the jet flare in the decaying hard state has a similar light curve morphology to a simultaneous X-ray flare (Section 3.3; Fig. 4b). In Fig. 5 we show the optical excess flux (the flare) versus the X-ray count rate. Though the optical excess flux is not significantly correlated with the X-ray count rate (the Spearman rank coefficient of $0.79 \pm 0.11$ implies a significance of $1.1 \pm 0.3$ $\sigma$ for the eight data points), it can be fitted with the relation $F_{\rm \nu,optical~excess} \propto F_{\rm X}^{0.72 \pm 0.12}$ (Fig. 5). The apparent correlation holds in the main flare observed after MJD 55340. Before this date the X-ray is still fading during the main part of the outburst, and the optical excess is not well correlated with X-ray.

If the optical excess originates in jet emission with a flat spectrum (as tentatively suggested above) in the self-absorbed regime that extends to radio frequencies, we would expect it to be correlated with the X-ray flux with a relation $F_{\rm \nu,optical,jet} \propto F_{\rm X}^{\sim 0.7}$ \citep{falcbi95,fend01,market03,heinsu03,gallet03,gallet06,kordet06}. This relation is expected both if the X-ray emission originates in the inflow, or if it is synchrotron emission from the outflowing jet. The observed relation of $F_{\rm \nu,optical~excess} \propto F_{\rm X}^{0.72 \pm 0.12}$ is closer to this value than two alternatives: $F_{\rm \nu,optical~excess} \propto F_{\rm X}$, which is expected in the case of optically thin synchrotron in both optical and X-ray regimes, and $F_{\rm \nu,optical~excess} \propto F_{\rm X}^{\leq 0.5}$, as may be expected from an accretion disc \citep[both the underlying viscous disc and the irradiated disc;][]{vanpet94,russet06}. We caution that the slope of the correlation is sensitive to the flux of the optical excess at low levels at the end of the flare, which assumes the extrapolation of the exponential decay in Fig. 4 is valid.

The total observed optical emission is shown to correlate with the X-ray emission (Fig. 6). Here we extend the optical--X-ray correlations seen by C11 (using \textit{Swift} XRT and UVOT data) to lower luminosities with the FT data. The optical flux in each filter is correlated with the X-ray flux with a relation $F_{\rm \nu,optical} \propto F_{\rm X}^{0.2-0.5}$. There is a discrepancy between the UVOT and FT optical fluxes; see Section 3.1). The shallowest relation is observed in the redder bands; $F_{\rm \nu,optical} \propto F_{\rm X}^{0.24 \pm 0.04}$ in both R and i$^{\prime}$-bands. The steepest relation from FT data is seen in B-band; $F_{\rm \nu,optical} \propto F_{\rm X}^{0.35 \pm 0.03}$ and the V-band lies between this and the redder bands. The UVOT relations, which were mainly during the soft state, are steeper; $F_{\rm \nu,optical} \propto F_{\rm X}^{0.4-0.5}$.

These relations are generally shallower than those expected from an irradiated disc; $F_{\rm \nu,optical} \propto F_{\rm X}^{0.5}$ \citep{vanpet94}. For a non-irradiated, viscously heated disc we expect correlation slopes on the order of what is observed, and shallower slopes for the redder bands, also as observed \citep*{franet02,russet06}. However, since the jet likely made a contribution in the fading hard state, this emission increased the optical flux, making the correlation appear shallower than if the disc alone was producing the emission. Subtracting the emission from the excess, we estimate correlations of $F_{\rm \nu,V} \propto F_{\rm X}^{0.39 \pm 0.05}$, $F_{\rm \nu,R} \propto F_{\rm X}^{0.40 \pm 0.06}$ and $F_{\rm \nu,i^{\prime}} \propto F_{\rm X}^{0.42 \pm 0.06}$ for V, R and i$^{\prime}$, respectively. These slopes are now marginally consistent with an irradiated disc, but scatter in the correlations is introduced because while the optical (non-excess) disc was supposedly fading, the X-ray flux was increasing then fading in the aforementioned flare. We can therefore make no firm conclusions from the correlations in Fig. 6, except that the optical emission most likely originates in the accretion disc.

\subsection{Constraints on the jet spectral break and the origin of the X-ray flare}

The PCA X-ray spectra are well fit by a broken power law in the fading hard state (Section 2.5; Fig. 4e) except after MJD 55355, when a single power law provides a better approximation (the counts are also low in most of these late observations). The two power laws have photon indices $\Gamma_1 \sim 2.0 \pm 0.2$ ($\alpha_1 \sim -1.0 \pm 0.2$) and $\Gamma_2 \sim 1.5 \pm 0.2$ ($\alpha_2 \sim -0.5 \pm 0.2$) in the majority of the hard state \citep[see also][]{stieet11}. The spectral index between the optical excess and the X-ray 3--20 keV flux is $\alpha = -0.51$ at the peak of the flare (around MJD $\sim 55360$). For optically thin synchrotron emission we expect $-0.8 \simlt \alpha \simlt -0.5$ depending on the lepton energy distribution in the jet. If both X-ray flux and optical excess originate in the jet, the measured spectral indices at optical ($\alpha = -0.16 \pm 0.29$), and X-ray energies and between the two ($\alpha = -0.51$) suggest an optically thick-to-thin spectral break in the UV, blueward of V-band. The optical spectral index is poorly constrained however, and appears to be optically thin near the end of the flare (Section 3.3). Alternatively, the X-ray flux may not be dominated by the jet emission, and the high energy cut-off in the jet spectrum may lie below 3 keV, outside the RXTE range.

The synchrotron jet cannot produce the majority of the X-ray flux before MJD $\sim$ 55340 (before the X-ray flare) because the optical jet emission is faint before that date, whereas the X-ray flux is bright (Fig. 4b). Instead, the X-ray power law probably originates in the Compton upscattering of soft photons on hot electrons in the corona before MJD $\sim$ 55340. After this date, the X-ray flare is contemporaneous with the optical jet flare, and a separate power law from the jet may explain the 40 day flare, becoming brighter than the corona power law.

We see from Fig. 4d and 4e that no abrupt changes in the X-ray rms variability amplitude, nor the X-ray spectrum are seen around MJD 55340. There is a clear increase in the rms variability over the state transition, well before the jet flare. There is an intriguing jump in the rms, from $\sim 20$ per cent to $\sim 26$ per cent around MJD 55325, during the hard state decay but before the flare. There are some apparent variations of the power law indices but no clear change around MJD 55340. There is also no obvious change in the energy of the break between the two power laws, which appears to be fit well using a break energy of $\sim 6$--10 keV before and during the flare.

We also binned the PCA data over a longer time period to gain signal-to-noise and decrease errors in the fit parameters. We combined data in the intervals MJD 55321 -- 55334 (before the flare) and MJD 55352 -- 55356 (during the flare, at similar flux levels). The resulting power law fits (shown as black circles and green squares in Fig. 4e) are very similar; $\Gamma_1 = 2.00^{+0.04}_{-0.05}$; $\Gamma_2 = 1.51^{+0.02}_{-0.03}$; break energy $\sim 6.5$ keV for the first interval, and $\Gamma_1 = 2.00^{+0.06}_{-0.08}$; $\Gamma_2 = 1.48^{+0.06}_{-0.05}$; break energy $\sim 6.6$ keV for the second interval. Either the corona and jet both have very similar X-ray properties, or the X-ray corona and optical jet brightened and faded simultaneously.

It was shown by \cite{caseet10} that IR light from the jet of GX 339--4 was positively correlated with the X-ray flux, with an IR lag of 100 ms with respect to X-ray (from simultaneous observations made in the hard state). Perturbations in the accretion flow are likely to affect the jet variability, so it is not surprising that the X-ray properties of the corona (if the X-rays originate in the corona) and jet are similar. If the observational differences between jet and corona are subtle, it is difficult to distinguish between them without multi-wavelength data. One way to differentiate between the two may be X-ray polarization. The jet will be strongly polarized if the poloidal magnetic field in the emitting region is fairly ordered (see Section 3.7). Conversely, the corona is likely to be more chaotic, with no strong net magnetic field orientation.

In the X-ray jet scenario, the appearance of the jet in the fading hard state is actually more prominent at X-ray energies than it was for XTE J1550--564 because the X-ray flux is seen as a flare, increasing and decreasing in intensity. With XTE J1550--564, the flare was visible as an excess above the decaying corona component. However, in XTE J1550--564 the case for the jet dominating the X-ray luminosity was strengthened by the linear correlation between OIR jet and X-ray flux, which favoured a single power law producing the whole OIR--X-ray spectrum. Here, the optical--X-ray data of \1752 favour a broken power law. Whatever the origin of the X-ray flare, it correlates well with the jet emitting in the optical.

Any constraint on the position of the optically thick-to-thin jet spectral break would benefit our understanding of jet physics, since the break directly relates to the magnetic field in the flow and the total radiative jet power  \citep*{miglet06,chatet11}. A jet spectral break in the UV would imply a stronger jet than commonly assumed in BHXBs; constraints currently place the jet break in the infrared \citep{miglet10,rahoet11,gandet11}, however breaks in the optical are favoured when modelling the broadband SEDs of some BHXBs \citep[e.g.][]{miglet07}. Robust estimates of the total luminosity of the jet are also difficult to achieve due to the uncertain position of the high energy cooling break in the jet spectrum. Here, the jet break in \1752 resides in the IR at the end of the jet flare since we see optically thin emission in the OIR. Before then, the break is unconstrained but weak evidence suggests it moves to spectral regimes as high as optical or UV.

\begin{figure}
\centering
\includegraphics[width=5.8cm,angle=270]{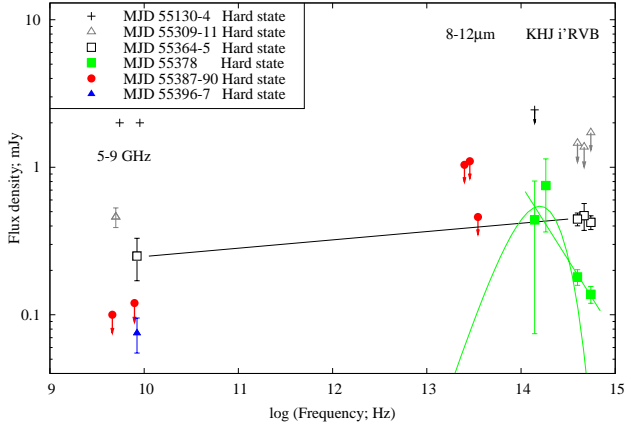}
\caption{Broadband, unabsorbed radio to optical SEDs of the isolated jet component (see text for details and assumptions). Symbols are the same as those used in Fig. 3.}
\end{figure}

\subsection{Spectral energy distributions of the jet}

In Section 3.3 we showed that the optical flare was due to jet emission, and we isolated the jet and disc emission during the hard state outburst decay. Since radio observations were made during this time \citep{brocet09,yanget10,millet11}, we can construct SEDs of the jet component from radio to optical, which are presented in Fig. 7. The jet was detected at radio frequencies in the hard state rise before the state transitions, but not unambiguously at OIR frequencies (black crosses in Fig. 7); the radio to NIR jet spectral index is $\alpha \leq +0.02$ at this time.

One radio detection of $0.25 \pm 0.08$ mJy at 8.4 GHz was made near the peak of the optical flare \citep{millet11}, and we constrain the radio to optical jet spectral index to be $\alpha = +0.05^{+0.05}_{-0.03}$ during the flare. The optical (i$^{\prime}$ to V-band) spectral index of the flare is $\alpha = -0.16 \pm 0.29$ (Section 3.3), consistent with the radio to optical spectrum within errors; further evidence that the optically thick (self absorbed) jet spectrum may extend even into the UV. The extrapolation of the X-ray power law to lower frequencies also favours a jet spectral break in the UV. However, the NIR to optical spectral index of the jet near the end of the flare is $\alpha = -1.0 \pm 0.3$ (green power law in Fig. 7), implying a dramatic change in the jet spectrum as the flare fades. A blackbody with a temperature of 2700 K (if main sequence this would be an M class star) is shown in the SED (green curve in Fig. 7) that may approximate the NIR--optical non-disc emission, but the companion cannot account for a large fraction of the NIR emission (Section 3.3). Instead, the spectral index of $\alpha = -1.0 \pm 0.3$ is consistent with optically thin synchrotron emission.

\subsection{Polarization of the jet}

Optically thin synchrotron emission from X-ray binary jets can be polarized, depending on the magnetic field configuration in the emitting region. Observationally, radio emission can be linearly polarized by tens of per cent, possibly from the formation of shocks downstream in the jets \citep[e.g.][]{fendet02,brocet07}. The compact, core jet emits optically thin synchrotron emission at OIR frequencies during hard states, and this may be polarized at a level of a few per cent \citep[as found in some XBs; e.g.][]{shahet08,russfe08}. The SEDs in Fig. 7 suggest this optically thin jet emission dominates the infrared emission during a time in the hard state in which we measure a linear polarization of $P_{\rm H} < 30$ per cent in both H and K$_{\rm S}$-bands. This implies the magnetic field near the base of the jet is at least moderately tangled \citep[a highly ordered field would give up to 70 per cent polarization for optically thin synchrotron emission;][]{rybili79,bjorbl82}. This is consistent with the fairly tangled (and sometimes variable) magnetic fields reported in other X-ray binaries from OIR linear polarization measurements \citep{shahet08,russfe08,russet11a} but differs from the highly ordered magnetic field recently claimed from $\gamma$-ray polarization of Cygnus X--1 \citep{lauret11}.

\subsection{Orbital period constraints}

The quiescent magnitudes of \1752 are B $\geq 20.6$, V $\geq 21.1$, R $\geq 19.5$, i$^{\prime}$ $\geq 19.2$. Using the empirical relation between V-band outburst amplitude (which is $> 4.1$ mag) and orbital period $P$ of \citeauthor{shahku98} (1998; which was found to exist for $P < 1$ d), we estimate the orbital period of \1752 is likely to be $< 22.1$ h (unless the system has a high inclination to the line of sight, in which case the projected surface area of the disc may be smaller so the disc would appear fainter). At $3.5 \pm 0.4$ kpc \citep[as suggested by][]{shapet10}, the quiescent absolute magnitude is $M_{\rm V} > 5.3$, which implies the donor star may be later than G5V \citep{cox00}, also consistent with a short orbital period. Alternatively the system may harbour a faint subgiant companion, as suggested for other BHXBs \citep*{king93,munoet08}. We caution that the spectral type estimate assumes the distance, orbital period and extinction are all correct within errors. In addition, according to \cite{laso08}, the X-ray luminosity of \1752 is expected to be low in quiescence ($L_{\rm X} \leq 3 \times 10^{31}$ erg s$^{-1}$; 0.5--10 keV) given its short orbital period.

We also remark that if the (possibly irradiated) companion star makes any contribution to the optical emission during the outburst decay, it could introduce periodic variability on a timescale of the orbital period. We cannot test for a period from these data however, due to the complexity of the long-term variability trends, uncertain rapid variability from the irradiated disc and/or the jet, and signal-to-noise ratio limitations.

\section{Summary}

We have presented multi-wavelength observations of the BHXB \1752 during its 2009--2010 outburst. Monitoring of the outburst decay in several optical filters revealed a slowly fading optical counterpart, which at late times dropped more rapidly towards quiescence. We are able to separate the disc and jet emission during the hard state decay. A blue optical spectrum, exponentially decaying in four filters can be described by a (possibly irradiated) accretion disc.

A redder 40 day optical flare is found to be most likely associated with synchrotron emission from the jet, and is contemporaneous with an X-ray flare of very similar light curve morphology, suggesting a common origin. Either the X-ray flare also originates in the jet, or the corona and jet are closely correlated during the flare. If the jet produces the X-ray flare, the jet and corona both have close to identical X-ray timing and spectral properties.

We also present the first mid-IR observations and the first NIR polarimetry of \1752, near the end of the jet flare. The NIR flux is not strongly linearly polarized (we measure $< 30$ per cent). If the synchrotron emission is optically thin (as implied by the SEDs), this suggests a magnetic field that is not highly ordered in the inner regions of the jet close to the black hole (in the optically thick regime the polarization would be reduced). The companion star may make a low level contribution to the NIR flux. The radio to optical jet spectrum is approximately flat. The jet break between optically thick and optically thin synchrotron emission resides in the IR near the end of the flare, but there is some evidence to suggest this break may shift to higher frequencies (optical or UV) at the peak of the jet flare (higher frequencies than usually found). This may result from a changing jet power, that increases then decreases as the jet brightens and fades.

The quiescent magnitudes of the optical counterpart are B $\geq 20.6$, V $\geq 21.1$, R $\geq 19.5$, i$^{\prime}$ $\geq 19.2$. From the optical outburst amplitude we estimate a likely orbital period of $< 22.1$ h.

\section*{Acknowledgments}

DMR would like to thank Mario van den Ancker at ESO for help with the preparation of VLT/VISIR observations, and Sam Rix, Ovidiu Vaduvescu and ING staff for their support at the WHT, in particular for extensive help with the LIRIS polarimetry mode. We thank the anonymous referee for suggesting additional analysis that proved very useful. \textit{Swift}/BAT transient monitor results are provided by the \textit{Swift}/BAT team. The Faulkes Telescopes (North and South) are maintained and operated by Las Cumbres Observatory Global Telescope Network. 
DMR and SM acknowledge support from a Netherlands Organisation for Scientific Research (NWO) Veni and Vidi Fellowship, respectively. PAC is supported by the Centre National d'Etudes Spatiales (CNES), France through MINE: the Multi-wavelength INTEGRAL NEtwork. FL acknowledges support from the Dill Faulkes Educational Trust. TB acknowledges support from ASI through contract ASI/INAF I/009/10/0. PC acknowledges funding via a EU Marie Curie Intra-European Fellowship under contract no. 2009-237722. The research leading to these results has received funding from the European Community's Seventh Framework Programme (FP7/2007-2013) under grant agreement number ITN 215212 `Black Hole Universe'.

\label{lastpage}

\end{document}